\documentclass[twocolumn,prb,showpacs]{revtex4}
\usepackage{graphicx}
\usepackage{dcolumn}
\usepackage{bm}
\begin{document}

\title{Effect of the Spatial Dispersion on the Shape of a Light Pulse in a Quantum Well}
\author{L. I. Korovin, I. G. Lang}
\address{A. F. Ioffe Physical-Technical Institute, Russian
Academy of Sciences, 194021 St. Petersburg, Russia}
\author{S. T. Pavlov\dag\ddag}
\address{\dag Facultad de Fisica de la UAZ, Apartado Postal C-580,
98060 Zacatecas, Zac., Mexico;}
\address{\ddag
P.N. Lebedev Physical Institute, Russian Academy of Sciences,
119991 Moscow, Russia; pavlov@sci.lebedev.ru}

\begin{abstract}
Reflectance, transmittance and absorbance of a symmetric light
pulse, the carrying frequency of which is close to the frequency
of interband transitions in a quantum well, are calculated. Energy
levels of the quantum well are assumed discrete, and two closely
located excited levels are taken into account. A wide quantum well
(the width of which is comparable to the length of the light wave,
corresponding to the pulse carrying frequency) is considered, and
the dependance of the interband matrix element of the momentum
operator on the light wave vector is taken into account.
Refractive indices of barriers and quantum well are assumed equal
each other. The problem is solved for an arbitrary ratio of
radiative and nonradiative lifetimes of electronic excitations. It
is shown that the spatial dispersion essentially affects the
shapes of reflected and transmitted pulses. The largest changes
occur when the radiative broadening is close to the difference of
frequencies of interband transitions taken into account.
\end{abstract}


\maketitle

Irradiation of the low-dimensional semiconductor systems by light
pulses and analysis of reflected and transmitted pulses allow to
obtain the information regarding the structure of energy levels as
well as relaxation processes.

The radiative mechanism of relaxation of excited energy levels in
quantum wells arises due to a violation of the translation
symmetry perpendicular to the he quantum well plane
\cite{bb1,bb2}. At low temperatures, low impurity doping and
perfect boundaries of quantum wells, the contributions of the
radiative and nonradiative relaxation can be comparable. In such
situation, one cannot be limited by the linear approximation on
the electron - light interaction. All the orders of the
interaction have to be taken into account
\cite{bb3,bb4,bb5,bb6,bb7,bb8,bb9}. Alterations of asymmetrical
\cite{bb10,bb11,bb12,bb13} and symmetrical \cite{bb13,bb14,bb15}
light pulses are valid for narrow quantum wells under conditions
$k\,d\ll\,1$ ($d$ is the quantum well width, $k$ is the magnitude
of the light wave vector corresponding to the carrying frequency
of the light pulse) and an independence of optical characteristics
of a quantum well on $d$. However, a situation is possible  when
the size quantization is preserved and for wide quantum wells if
$k\,d\geq\,1$ (see corresponding estimates in \cite{bb16}). In
such a case, we have to take into account the spatial dispersion
of a monochromatic wave $\cite{bb9,bb19}$ and waves composing the
light pulse \cite{bb16}.

Our investigation is devoted to the influence of the spatial
dispersion on the optical characteristics (reflectance,
transmittance and absorbance) of a quantum well irradiated by the
symmetric light pulse. A system, consisting of a deep quantum well
of type I, situated inside of the space interval $0\leq z\leq d$,
and two semi-infinite barriers, is considered. A constant
quantizing magnetic field is directed perpendicular to the quantum
well plane what provides the discrete energy levels of the
electron system. A stimulating light pulse propagates along the
$z$ axis from the side of negative values $z$. The barriers are
transparent for the light pulse which is absorbed in the quantum
well to initiate the direct interband transitions. The intrinsic
semiconductor and zero temperature are assumed.

The final results for two closely spaced energy levels of the
electronic system in a quantum well are obtained. Effect of other
levels on the optical characteristics may be neglected, if the
carrying frequency $\omega_{\,\ell}$ of the light pulse is close
to the frequencies $\omega_{\,1}$ and  $\omega_{\,2}$ of the
doublet levels, and other energy levels are fairly distant. It is
assumed that the doublet is situated near the minimum of the
conduction band, the energy levels may be considered in the
effective mass approximation, and the barriers are infinitely
high.

In the case $\hbar\,K_\perp =0$ ($\hbar\,K_\perp$ is the vector of
the quasi-momentum of electron-hole pair in the quantum well
plane) in a quantum well, the discrete energy levels are the
excitonic energy levels in a zero magnetic field or energy levels
in a quantizing magnetic field directed perpendicularly to the
quantum well plane. As an example, the energy level of the
electron-hole pair in a quantizing magnetic field directed along
the $z$ axis (without taking into account the Coulomb interaction
between the electron and hole which is a weak perturbation for the
strong magnetic fields and not too wide quantum wells) is
considered.

\section{The electric field}

Let us consider a situation when a symmetric exciting light pulse
propagates through a single quantum well along the $z$ axis from
the side of negative values of $z$. Analogously to
\cite{bb13,bb14,bb15}, the electric field is chosen as
\begin{eqnarray}
\label{eq1} &&{\bf E}_{\,0}\,(z,t)={\bf
e}_{\,\ell\,}E_{\,0\,}e^{-i\omega_{\,\ell\,}p}\nonumber\\&&\times
\left\{\Theta(p)e^{-\gamma_{\,\ell\,}p/2}\,+
\,[1-\Theta(p)]e^{\gamma_{\,\ell\,}p/2}\right\}+c.c.,
\end{eqnarray}
where $E_{\,0}$ is the real amplitude, $p=t-\nu\,z/\,c$,
$${\bf e}_{\,\ell}={1\over\sqrt{2}}({\bf e}_{\,x}\,\pm\,i\,{\bf e}_{\,y})$$
are the unite vectors of the circular polarization, ${\bf
e}_{\,x}$ и ${\bf e}_{\,y}$ are the real unite vectors,
$\Theta(p)$ is the Heaviside function, $1/\gamma_{\,\ell}$
determines the pulse width, $c$ is the light velocity in vacuum,
$\nu$ is the refraction index, which is assumed the same for the
quantum well and barriers (the approximation of a homogeneous
media). The Fourier-transform of (\ref{eq1}) is as follows
\begin{eqnarray}
\label{eq2}
 {\bf E}_{\,0}(z,\omega )\,=\,e^{ikz}\left[{\bf
e}_{\,\ell}\,E_{\,0}(\omega)+
{\bf e}_{\,\ell}^*\,E_{\,0}(-\omega )\right],\nonumber\\
E_{\,0}(\omega)\,=\,\frac{E_{\,0}\,\gamma_{\,\ell}}{(\omega-\omega_
{\,\ell})^2+(\gamma_{\,\ell}/2)^2},~~~~k\,=\,\frac{\nu\omega}{c}.
\end{eqnarray}

The electric field in the region $z\leq 0$ consists of the sum of
the exciting and reflected pulses. The Fourier-transform may be
written as
$${\bf E}^{\,\ell}\,(z,\omega)\,=\,{\bf E}_{\,0}\,(z,\omega)\,+
\,\Delta{\bf E}^{\,\ell}\,(z,\omega),$$ where $\Delta{\bf
E}^{\,\ell}(z,\omega)$ is the electric field of the reflected
pulse
\begin{equation}
\label{eq3} \Delta{\bf E}^{\,\ell}(z,\omega)\,= \,{\bf
e}_{\,\ell}\,\Delta E^{\,\ell}(z,\omega)\,+ \,{\bf
e}_{\,\ell}^{\,*\,}\,\Delta E^{\,\ell}(z,-\omega).
\end{equation}
In the region $z\geq d$, there is only the transmitted pulse, and
its electric field is
\begin{equation}
\label{eq4} {\bf E}^{\,r}(z,\omega)\,= \,{\bf e}_{\,\ell}\,
E^{\,r}(z,\omega)\,+ \,{\bf e}_{\,\ell}^{\,*\,}\,
E^{\,r}(z,-\omega).
\end{equation}

It is assumed below that the  pulse, having absorbed in the
quantum well, stimulates the interband transitions and,
consequently, the appearance of a current. In barriers, the
absorption is absent. Therefore, for the complex amplitudes
$\Delta E^{\,\ell}(z,\omega)$ and $E^{\,r}(z,\omega)$ in barriers
for $z\leq 0$ and $z\geq d$, we obtain the expression
\begin{equation}
\label{eq5} \frac{d^{\,2}E}{d\,z^{\,2}}\,+\,k^{\,2}\,E\,=\,0.
\end{equation}
The expression for the electric field inside of the quantum well
$(0\leq z \leq d)$ has a form
\begin{equation}
\label{eq6} \frac{d^{\,2}E}{d\,z^{\,2}}\,+\,k^{\,2}\,E\,=
\,-\frac{4\pi i\,\omega}{c^{\,2}}\,J(z,\omega),
\end{equation}
where $J(z,\omega)$ is the Fourier-transform of the current
density, averaged on the ground state of the system. The current
is induced by the monochromatic wave of the frequency $\omega$. In
the case of two excited energy levels, $J(z,\omega)$ is expressed
as follows
\begin{equation}
\label{eq7} J(z,\omega)\,=\,\frac{i\nu c}{4\pi}\sum_{j=1}^2\,
\frac{\gamma_{\,rj}\,\Phi_{\,j}(z)}{\tilde {\omega}_{\,j}}
\int_0^d\,dz'\,\Phi_{\,j}(z')\,E(z^{\,\prime},\omega),
\end{equation}
\begin{equation}
\label{eq8}
\tilde{\omega}_{\,j}\,=\,\omega\,-\,\omega_{\,j}\,+\,i\gamma_{\,j}/2.
\end{equation}
where $\gamma_{\,j}$ is the nonradiative damping of the doublet,
$\gamma_{\,r\,j}$ is the radiative damping of the levels of the
doublet in the case of narrow quantum wells, when the spatial
dispersion of electromagnetic waves may be neglected.

In particular, the doublet system may be represented by a
magnetopolaron state \cite{bb18}. In such a case,
$$\gamma_{\,r,j}\,=\,\gamma_{\,r}\,Q_{\,j},~~~~~~
\gamma_{\,r}\,=\,\frac{2e^2}{\hbar\,c\nu}\,\frac{p^{\,2}_{\,c\,v}}
{\hbar\,\omega_{\,g}\,m_{\,0}}\,\frac{|e|H}{m_{\,0}\,c},$$ where
$m_{\,0}$ is the free electron mass, $H$ is the magnetic field,
$e$ is the electron charge, $p_{\,c\,v\,}$ is the matrix element
of the momentum, corresponding to the circular polarization,
$p_{\,c\,v\,}^2=|p_{\,c\,v\,x\,}|^2+ |p_{\,c\,v\,y\,}|^2$. The
factor
$$Q_{\,j} = {1\over 2}\pm{\hbar\,(\Omega_{\,c} - \omega_{\,LO})\over
2\sqrt{\hbar^2(\Omega_{\,c} - \omega_{\,LO})^2+ (\Delta
E_{\,pol})^2}} $$ determines the change of the radiative timelife
at a deflection of the magnetic field from the resonant value when
the resonant condition $\Omega_{\,c}\,=\,\omega_{\,LO}$ is carried
out. $\Delta E_{\,pol}$ is the polaron splitting, $\Omega_{\,c}$
and $\omega_{\,LO}$ are the cyclotron  frequency and optical
phonon frequency, respectively. In the resonance,
 $Q_{\,j}=1/2$ and $\gamma_{\,r1}\,=\,\gamma_{\,r2}$.

When calculating $J(z,\omega)$, it was assumed that the Lorentz
force, determined by the external magnetic field, is large in
comparison with  the Coulomb and exchange forces in the
electron-hole pair. In that case, the variables $z$ (along
magnetic field) and ${\bf r}_\perp$ (in the quantum well plane) in
the wave function of the electron-hole pair may be separated. This
condition is carried out for the quantum well on basis of GaAs for
the magnetic field, corresponding to the magnetopolaron formation
\cite{bb9}. Besides, if the energy of the size quantization
exceeds the Coulomb and exchange energies,  the electron-hole pair
may be considered as a free particle. Then, in the approximation
of the effective mass and infinitely high barriers, the wave
function, describing the dependence on $z$ , accepts a simple form
\begin{equation}
\label{eq9} \Phi_{\,j}(z) = {2\over d}\,sin{\pi m_{\,c}z\over
d}\,sin{\pi m_{\,v}z\over d},~~ 0\leq z\leq d,
\end{equation}
and $\Phi_{\,j}(z)\,=\,0$ in barriers, where $m_{\,c}\,(m_{\,v})$
are the quantum numbers of the size-quantization of an electron
(hole).

In the real systems, the approximation (\ref{eq9}) is not always carried out.
However, the taking into account the Coulomb and exchange interactions will
result only into some changes of the function $\Phi_{\,j}(z)$, what does not
change qualitatively the optical characteristics, as it was shown for the
monochromatic irradiation \cite{bb20}.

Indices $j=1$ and  $j=2$ in $\Phi_{\,j}(z)$ correspond to the
pairs of quantum numbers of the size-quantization in a direct
interband transition. $m^{(1)}_{\,c\,(\,v)}$ corresponds to the
index $j=1$, and $m^{(2)}_{\,c\,(\,v)}$ corresponds to the index
$j=2$. In interband transitions, the Landau quantum numbers are
conserved. The total electric field $E$ is included into the RHS
of (\ref{eq7}), what is connected with the refuse from the
perturbation theory on the coupling constant $e^2/\hbar c$.

In further calculations, an equality of quantum numbers
$m^{(1)}_{\,c\,v}\,=\,m^{(2)}_{\,c\,v}$ is assumed. Then,
$$\Phi_{\,1}(z)\,=\,\Phi_{\,2}(z)\,=\,\Phi(z),$$
and the current density in the RHS of (\ref{eq7}) takes the form
\begin{eqnarray}
\label{10} &&J(z,\omega)\,=\,\frac{i\,\nu\,c}{4\,\pi}\bigg (\frac
{\gamma_{\,r1}}{{\omega}\,-\,\omega_{\,1}\,+\,i\gamma_{\,1}}
+\frac{\gamma_{\,r2}}{{\omega}-\,\omega_{\,2}\,
+\,i\gamma_{\,2}}\bigg
)\nonumber\\
&&\times\Phi(z)\int_0^ddz'\Phi(z')E(z').
\end{eqnarray}
With the help of the indicated simplifications, as it was shown in
\cite{bb9,bb18,bb20}, the field amplitudes in the
Fourier-representation $\Delta E^{\,\ell}(z,\omega)$ and
$E^{\,r}(z,\omega)$ result in
\begin{eqnarray}
\label{eq11} &&\Delta E^{\,\ell}(z,\omega)\,=\,
-iE_0(\omega)(-1)^{m_{\,c}+m_{\,v}}e^{-ik(z-d)}{\cal
N},\nonumber\\
&& E^{\,r}(z,\omega)\,=\,E_0(\omega)e^{ikz}(1-i{\cal N}),
\end{eqnarray}
where $E_0(\omega)$ is given in (\ref{eq3}). Here, the frequency
dependence is determined by the function
\begin{equation}
\label{eq12} {\cal
N}\,=\,\frac{\varepsilon\,(\gamma_{\,r1}\,\tilde{\omega}_2\,+\,
\gamma_{\,r2}\,\tilde{\omega}_1)/2}{\tilde{\omega}_1\,\tilde{\omega}_2\,+\,
i\varepsilon(\gamma_{\,r1}\,\tilde{\omega}_2\,+\,
\gamma_{\,r2}\,\tilde{\omega}_1)/2}.
\end{equation}

The function ${\cal N}$ includes the value
\begin{equation}
\label{eq13} \varepsilon\,=\,\varepsilon'\,+\,i\varepsilon'',
\end{equation}
%
 which
determines influence of the spatial dispersion on the radiative
broadening ($\varepsilon'\,\gamma_{\,r}$) and shift
($\varepsilon''\,\gamma_{\,r}$) of the doublet levels.
$\varepsilon'$ and $\varepsilon''$ are equal \cite{bb9,bb20}:
\begin{equation}
\label{eq14} \varepsilon'\,=\,Re\,\varepsilon\,=\,2{\cal B}^2\,
\big [1-(-1)^{m_{\,c}+m_{\,v}}\cos kd\,\big ],
\end{equation}
\begin{eqnarray}
\label{eq15} &&\varepsilon''=Im \varepsilon=2{\cal
B}\nonumber\\
&&\times\Bigg
(\frac{(1+ \delta_{m_{c},m_{v}})(m_{c}+m_{v})^2+(m_{c}-m_{v})^2}{8m_{c}m_{v}}\nonumber\\
&& -(-1)^{m_{c}+m_{v}}{\cal B}\sin kd-\frac{(2+
\delta_{m_{c},m_{v}})(kd)^2}{8\pi^2m_{c}m_{v}}\Bigg ),
\end{eqnarray}
$${\cal B}\,=\,\frac{4\pi^{\,2}\,m_{\,c}\,m_{\,v}\,kd}
{[\pi^{\,2}\,(m_{\,c}+m_{\,v})^2-(kd)^2\,]\,[(kd)^2\,-\pi^{\,2}\,(m_{\,c}-
m_{\,v})^2\,]}$$ (if $kd\to 0$, то $\varepsilon\to
1\,(m_{\,c}\,=\,m_{\,v}$, an allowed transition ) and
$\varepsilon\to 0\,(m_{\,c}\,\neq\,m_{\,v}$, a forbidden
transition)).

\section{The time-dependence of the electric field of reflected and transmitted
light pulses}

With the help of the standard formulas, let us go to the
time-representation
\begin{eqnarray}
\label{eq16}&& \Delta E^{\,\ell}(z,t)\,\equiv\,\Delta
E^{\,\ell}(s)\,=\nonumber\\&&=\frac{1}{2\pi}
\int_{-\infty}^{+\infty}d\omega\,e^{-i\omega\,s}\,\Delta
E^{\,\ell}(z,\omega),~~  s=t+\nu z/c,~~~~~
\end{eqnarray}
\begin{eqnarray}
\label{eq17} &&E^{\,r}(z,t)\,\equiv\,
E^{\,r}(p)\,=\,\frac{1}{2\pi}
\int_{-\infty}^{+\infty}d\omega\,e^{-i\omega\,p}\,
E^{\,r}(z,\omega), \nonumber\\ &&p=t-\nu z/c.
\end{eqnarray}
The vectors $ \Delta {\bf E}^{\,\ell}(s)$ and $ {\bf E}^{\,r}(p)$
have the form
\begin{eqnarray}
\label{eq18} &&\Delta {\bf E}^{\,\ell}(s)\,=\,{\bf
e}_{\,\ell}\,\Delta E^{\,\ell}(s)+c.c., \nonumber\\ &&{\bf
E}^{\,r}(p)\,=\,{\bf e}_{\,\ell}\, E^{\,r}(p)+c.c..
\end{eqnarray}
It is seen from expressions $(\ref{eq11})$ and $(\ref{eq12})$ that
the denominator in integrands of $(\ref{eq16})$ and $(\ref{eq17})$
is the same. It may be transformed conveniently to the form
\begin{eqnarray}
\label{eq19} \tilde {\omega}_1\,\tilde
{\omega}_2\,+\,i(\varepsilon/2)\, (\gamma_{\,r1}\,\tilde
{\omega}_2\,+\,\gamma_{\,r2}\,\tilde {\omega}_1)\nonumber\\
=(\omega-\Omega_{\,1})\,(\omega-\Omega_{\,2}),
\end{eqnarray}
where $\Omega_{\,1}$ and $\Omega_{\,2}$ determine the poles of the
integrand in the complex plane  $\omega$. They are equal
\begin{eqnarray}
\label{eq20} \Omega_{1,2}=\frac{1}{2}\Bigg\{
\omega_1+\omega_2-\frac{i}{2}(\gamma_{1}+\gamma_{2})-
\frac{i\varepsilon}{2}(\gamma_{r1}+\gamma_{r2})\nonumber\\
\pm\Big([\omega_1-\omega_2-{i\over 2}(\gamma_{1}-\gamma_{2})-
{i\varepsilon\over 2}(\gamma_{r1}-\gamma_{r2})]^2\nonumber\\
- \varepsilon^2\gamma_{r1}\gamma_{r2}\Big)^{1/2}\Bigg \}.
\end{eqnarray}
Thus, in the integrands of \ref{eq16}) and (\ref{eq17}), there are
4 poles: $\omega\,=\,\omega_{\,\ell}\pm i\gamma_{\,\ell}/2$ and
$\omega=\Omega_{\,1,2}$. The pole
$\omega\,=\,\omega_{\,\ell}+i\gamma_{\,\ell}/2$ is situated in the
upper half plane, others are  situated in the lower half plane.

Integrating in the complex plane  $\omega$, we obtain that the
function $\Delta E^{\,\ell}(z,t)$, determining, according to
(\ref{eq17}), the electric field vector of the reflected pulse
$\Delta{\bf E}^{\,\ell}(z,t)$, has the form
\begin{eqnarray}
\label{eq21} \Delta E^{\,\ell}(z,t)\,=\,-iE_{\,0}
(-1)^{m_{\,c}+m_{\,v}}e^{ikd}\nonumber\\\times
\{R_1[1-\Theta(s)]+(R_2+R_3+R_4)\Theta(s)\},
\end{eqnarray}
where
\begin{eqnarray}
\label{eq22}
&&R_1\,=\,\exp(-i\omega_{\,\ell}s+\gamma_{\,\ell}s/2)\nonumber\\
&&\times\Bigg
(\frac{\bar{\gamma}_{\,r1}/2}{\omega_{\,\ell}-\Omega_{1}+i\gamma_{\,\ell}/2}+
\frac{\bar{\gamma}_{\,r2}/2}{\omega_{\,\ell}-\Omega_{2}+
i\gamma_{\,\ell}/2}\Bigg ),\nonumber\\
&&R_2\,=\,\exp(-i\omega_{\,\ell}-\gamma_{\,\ell}s/2)\nonumber\\
&&\times\Bigg
(\frac{\bar{\gamma}_{\,r1}/2}{\omega_{\,\ell}-\Omega_{1}-i\gamma_{\,\ell}/2}+
\frac{\bar{\gamma}_{\,r2}/2}{\omega_{\,\ell}-\Omega_{2}-
i\gamma_{\,\ell}/2}\Bigg ),\nonumber\\
&&R_3\,=\,-\exp(-i\Omega_1s)(\bar{\gamma}_{\,r1}/2)\nonumber\\
&&\times\Bigg
(\frac{1}{\omega_{\,\ell}-\Omega_{1}-i\gamma_{\,\ell}/2}-
\frac{1}{\omega_{\,\ell}-\Omega_{1}+ i\gamma_{\,\ell}/2}\Bigg
),\nonumber\\
&&R_4\,=\,-\exp(-i\Omega_2s)(\bar{\gamma}_{\,r2}/2)\nonumber\\
&&\times\Bigg
(\frac{1}{\omega_{\,\ell}-\Omega_{2}-i\gamma_{\,\ell}/2}-
\frac{1}{\omega_{\,\ell}-\Omega_{2}+ i\gamma_{\,\ell}/2}\Bigg ),
\end{eqnarray}
where
\begin{eqnarray}
\label{eq23}&&\bar{\gamma}_{\,r1}\,=\,\varepsilon'\,\gamma_{\,r1}\,+\Delta\gamma,
\bar{\gamma}_{\,r2}\,=\,\varepsilon'\,\gamma_{\,r2}\,-\Delta\gamma,\nonumber\\
&&\Delta\gamma\,=\,\frac{\varepsilon'\,\gamma_{\,r1}(\Omega_2-\omega_2+
i\gamma_2/2)}{\Omega_1-\Omega_2}\nonumber\\
&&+\frac{\varepsilon'\,\gamma_{\,r2}(\Omega_1-\omega_1+i\gamma_1/2)}
{\Omega_1-\Omega_2}.
\end{eqnarray}

The function $E^{\,r}(z,t)$, corresponding to a transmitted light
pulse, is represented in the form
\begin{equation}
\label{eq24}
E^{\,r}(z,t)\,=E_{\,0}{T_1[1-\Theta(p)]+(T_2+T_3+T_4)\Theta(p)\over
\Omega_1-\Omega_2},
\end{equation}
where
\begin{eqnarray}
\label{eq25} &&T_1\,=\,\exp(-i\omega_{\,\ell}p+\gamma_{\,\ell}p/2)
M(\omega_{\,\ell}+i\gamma_{\,\ell}/2)\nonumber\\
&&\times \Bigg
(\frac{1}{\omega_{\,\ell}-\Omega_1+i\gamma_{\,\ell}/2}-
\frac{1}{\omega_{\,\ell}-\Omega_2+i\gamma_{\,\ell}/2}\Bigg
),\nonumber\\
&&T_2\,=\,\exp(-i\omega_{\,\ell}p-\gamma_{\,\ell}p/2)
M(\omega_{\,\ell}-i\gamma_{\,\ell}/2)\nonumber\\
&&\times \Bigg
(\frac{1}{\omega_{\,\ell}-\Omega_1-i\gamma_{\,\ell}/2}-
\frac{1}{\omega_{\,\ell}-\Omega_2-i\gamma_{\,\ell}/2}\Bigg
),\nonumber\\
&&T_3\,=\,-\exp(-i\Omega_1\,p) M(\Omega_1)\nonumber\\
&&\times \Bigg
(\frac{1}{\omega_{\,\ell}-\Omega_1-i\gamma_{\,\ell}/2}-
\frac{1}{\omega_{\,\ell}-\Omega_1+i\gamma_{\,\ell}/2}\Bigg
),\nonumber\\
&&T_4\,=\,\exp(-i\Omega_2\,p) M(\Omega_2)\nonumber\\
&&\times \Bigg
(\frac{1}{\omega_{\,\ell}-\Omega_2-i\gamma_{\,\ell}/2}-
\frac{1}{\omega_{\,\ell}-\Omega_1+\gamma_{\,\ell}/2}\Bigg ).
\end{eqnarray}
The function $M$ has the structure
$$M(\omega)=(\omega-\omega_1+i\gamma_1/2)(\omega-\omega_2+i\gamma_2/2)-$$
$$(\varepsilon''/2)[\gamma_{r1}(\omega-\omega_{2}+i\gamma_{2}/2)+\gamma_{r2}
(\omega-\omega_{1}+i\gamma_{1}/2)].$$
 When the  electric field of stimulating light pulse ${\bf E}^{\,0}(z,t)$
(determined in (\ref{eq2})) is extracted from ${\bf E}^{\,r}(z,t)$
, i.e., it is assumed
\begin{equation}
\label{eq26} {\bf E}^{\,r}(z,t)\,=\,{\bf E}^{\,0}(z,t)+\Delta{\bf
E}^{\,r}(z,t),
\end{equation}
then, $\Delta{\bf E}^{\,r}(z,t)$ will differ from $\Delta{\bf
E}^{\,\ell}(z,t)$ only by substitution of the variable $s=t+\nu
z/c$ by $p=t-\nu z/c$ and by absence of the factor
$(-1)^{m_{\,c}+m_{\,v}}\exp(ikd)$.
 \begin{figure}
 \includegraphics [] {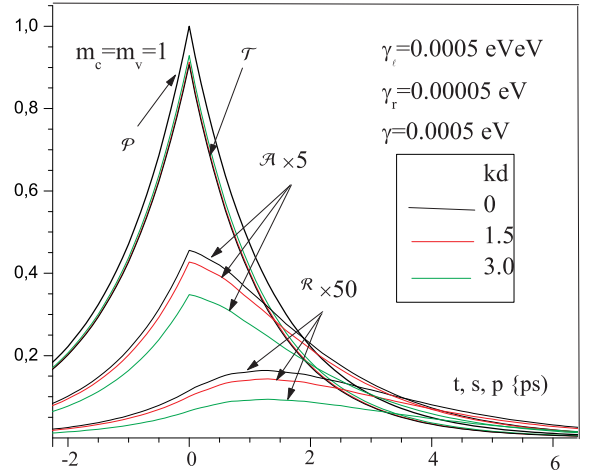} 
 \caption[*]{\label{Fig1.eps} The reflectance  ${\cal R}$,
 transmittance ${\cal T}$, absorbance ${\cal A}$, and
stimulating pulse ${\cal P}$ as time dependent functions for three
magnitudes of the parameter $kd$ in the case of a long stimulating
pulse  $(\gamma_{\,\ell\,}\ll \Delta \omega)$ и $\gamma_{\,r}\ll
\gamma, \gamma_{\,\ell\,}.\, \Delta \omega = 6.65.10^{-3}
eV,\,\omega_{\,\ell\,} = Re\,\Omega_1 = \Omega_{\,res}.$}
 \end{figure}
 \begin{figure}
 \includegraphics [] {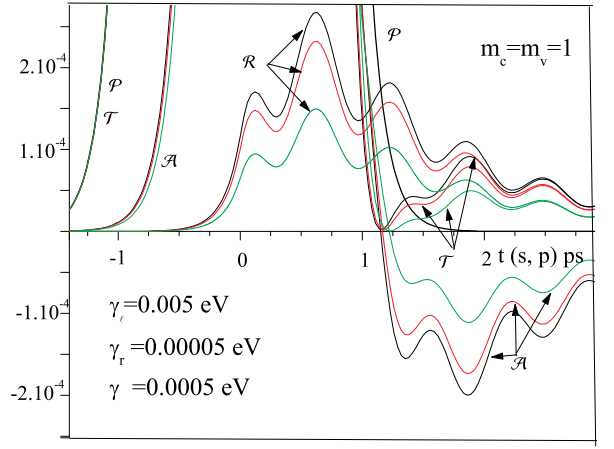} 
 \caption[*]{\label{Fig2.eps} Same as in Fig.1 for an exciting pulse of a middle duration
  $(\gamma_{\,\ell\,}\simeq \Delta
\omega)$ и $\gamma_{\,r}\ll \gamma \ll \gamma_{\,\ell\,}.$}
 \end{figure}

Thus, being taken into account, the spatial dispersion provides a
renormalization of radiative damping $\gamma_{\,r\,i}$. In
nominators of formulas (\ref{eq21}), the renormalization leads to
multiplication of  $\gamma_{\,r\,i}$ on the real factor
$\varepsilon '$, i.e., decreases the value $\gamma_{\,r\,i}$
(diagrams of functions $\varepsilon '$ and $\varepsilon ''$ are
represented in \cite{bb9}). In denominators, $\gamma_{\,r\,i}$ is
multiplied on the complex function
 $\varepsilon $, that means the appearance, together with the change of the radiative
 broadening, of a shift of resonant frequencies. In the limit $kd=0$,
expressions (\ref{eq21}) - (\ref{eq23}) coincide with obtained in
(\ref{eq14}).

\section{The reflectance, transmittance and absorbance of stimulating light pulse}

The energy flux ${\bf S}(p)$, corresponding to the electric field
of stimulating light pulse, is equal
\begin{equation}
\label{eq27} {\bf S}(p)\,=\frac{\mathbf{e_{z}}}{4\pi}\frac{c}{\nu}({\bf
E}^{\,0}(z,t))^2\,=\, {\bf e}_zS_0{\cal P}(p),
\end{equation}
where $S_0=cE_0^2/2\pi\nu, {\bf e}_z$ is the unite vector along
the light pulse. The dimensionless function
\begin{equation}
\label{eq28} {\cal P}(p)={({\bf E}^{\,0}(z,t))^2\over S_0}=
\Theta(p)e^{-\gamma_{\ell}p}+[1-\Theta(p)]e^{\gamma_{\ell}p}
\end{equation}
determines the spatial and time dependence of the energy flux of
stimulating pulse. The flux, transmitted through the quantum well,
has a form
\begin{equation}
\label{eq29} {\bf S}^r\,={{\bf e}_zc\over 4\pi\nu}({\bf
E}^{\,r}(z,t))^2\,=\, {\bf e}_zS_0{\cal T}(p),
\end{equation}
the reflected energy flux has a form
\begin{equation}
\label{eq30} {\bf S}^{\,\ell}\,=-{{\bf e}_zc\over 4\pi\nu}({\bf
E}^{\,\ell}(z,t))^2\,=\, -{\bf e}_zS_0{\cal R}(s).
\end{equation}
The dimensionless functions ${\cal T}(p)$ and ${\cal R}(s)$
correspond to parts of transmitted and reflected energy  fluxes of
the stimulating pulse. The dimensionless absorbance is defined as
\begin{equation}
\label{eq31} {\cal A}(p)\,=\,{\cal P}(p)-{\cal R}(p)-{\cal T}(p)
\end{equation}
(since for reflection $z\leq 0$, the variable in ${\cal R}$ is
$s=t-|z|/c$).

The dependencies of the reflectance  ${\cal R}$, transmittance
${\cal T}$, absorbance ${\cal A}$, and stimulated momentum ${\cal
P}$ on the variable  $p$ (or $s$ for ${\cal R}$) for the case
$m_{\,c\,}=m_{\,v\,}=1$ are represented in figures. It was assumed
also that
\begin{equation}
\label{eq32}
\gamma_{\,r\,1}\,=\,\gamma_{\,r\,2}\,=\,\gamma_{\,r},~~~~
\gamma_{\,1}=\gamma_{\,2}=\gamma.
\end{equation}
It follows from (\ref{eq21}) and (\ref{eq24}) that the resonant
frequencies are $\omega_{\,\ell}=Re\,\Omega_1$ and
$\omega_{\,\ell}=Re\,\Omega_2$. The calculations were performed
for
\begin{equation}
\label{eq33} \omega_{\,\ell}\,=\,Re\,\Omega_1\,=\,\Omega_{res}.
\end{equation}
Let us go from the frequency $\omega_{\,\ell}$ to
\begin{equation}
\label{eq34} \Omega\,=\,\omega_{\,\ell} -\omega_1,
\end{equation}
then the resonant frequency  is
\begin{equation}
\label{eq35} \Omega_{res}\,=\,\frac{1}{2}\big
[-\Delta\omega+\varepsilon'\gamma_{\,r}+
Re\,\sqrt{(\Delta\omega)^2-\varepsilon^2\gamma_{\,r}^2}\, \big ].
\end{equation}
It depends on three parameters: $\Delta\omega=\omega_1-\omega_2,
 \gamma_{\,r}$ and $kd$, since the complex function
$\varepsilon$ depends on $kd$ (see (\ref{eq15})).

Functions ${\cal R}, {\cal T}, {\cal A}$ and ${\cal P}$ are
homogeneous functions of the inverse lifetimes and frequencies
$\omega_1,
 \omega_2, \omega_{\,\ell}$. Therefore, a choice of the measurement units is arbitrary.
 For the sake of certainty, all these values are expressed in $eV$.
The time dependence of the optical characteristics of a quantum
well is represented in figures for the different magnitudes of
$kd$. The curves, corresponding to $kd=0$, were obtained in
\cite{bb14}. It was assumed in calculations that
$\Delta\omega=0.065eV$, what corresponds to the magnetopolaron
state in a quantum well on basis of GaAs and to the width
$d=300\AA$ of the  quantum well \cite{bb18,bb19,bb21}.

\section{The Discussion of results}

 Fig.1 corresponds to a long (wide in comparison to $\Delta\omega$)
stimulating pulse and a small radiative broadening
$(\gamma_{\,r}\ll\gamma,\gamma_{\,\ell})$. In this case, the transmittance
${\cal T}$ dominates. The shape of the curve weakly differs from  ${\cal P}$
and weakly depends on $kd$. The dependence on the spatial dispersion is seen
at the curves ${\cal R}$ and ${\cal A}$. For example, the reflectance  ${\cal
R}$ at $kd=3$ is two times less than at $kd=0$. However, the magnitude
 ${\cal R}$ is shares of percent.
 \begin{figure}
 \includegraphics [] {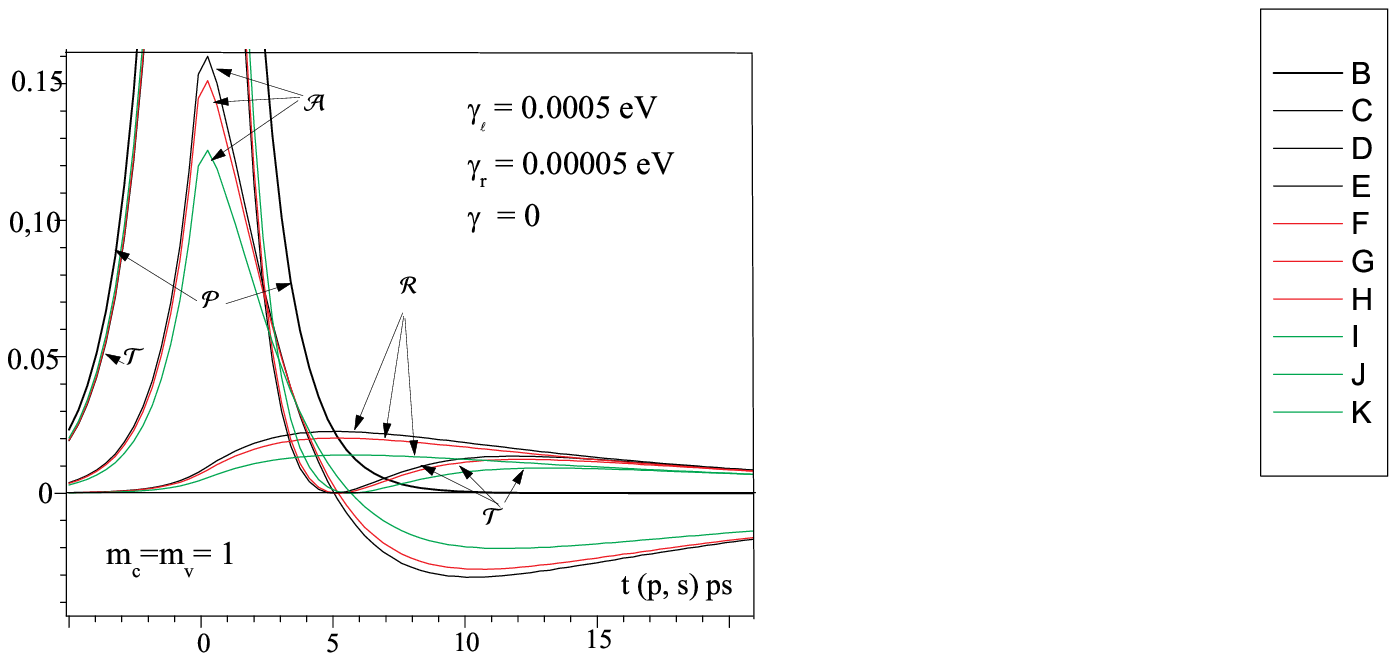} 
 \caption[*]{\label{Fig3.eps} Same as in Fig.1 for an exciting pulse of a middle duration
 $(\gamma_{\,\ell\,}\simeq \Delta
\omega)$ и $\gamma_{\,r}\ll \gamma \ll \gamma_{\,\ell\,}.$}
\end{figure}
 \begin{figure}
\includegraphics [] {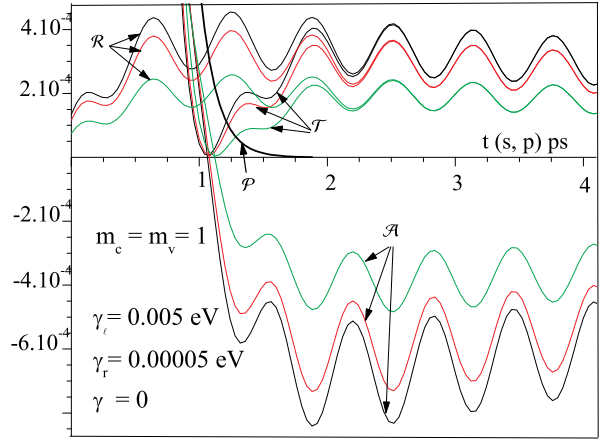} 
\caption[*]{\label{Fig4.eps} Same as in Fig.1 for an exciting
pulse of a middle duration $(\gamma_{\,\ell\,}\simeq \Delta
\omega)$ и $\gamma = 0.$}
\end{figure}

Fig.2 corresponds to a stimulating light pulse of a middle length,
when $\gamma_{\,\ell}\simeq\Delta\omega$ and
$\gamma_{\,r}\ll\gamma\ll\gamma_{\,\ell}$. There appear
peculiarities: a light generation (negative absorption) after the
light pulse transmission and oscillations of ${\cal R}, {\cal A} $
and ${\cal T}$. The generation is a consequence of the fact that
the electronic system has no time to irradiate the energy during
the propagation of such pulse. Oscillations is a consequence  of
beatings with the frequency (under condition
$\omega_{\,\ell}=Re\,\Omega_1$)
\begin{equation}
\label{eq36} Re\,(\Omega_1\,-\,\Omega_2)\,=
\,Re\,\sqrt{(\omega_1-\omega_2)^2-(\varepsilon'+i\varepsilon'')^2\gamma_{\,r}^2}.
\end{equation}
A noticeable effect of the spatial dispersion takes place in the
reflectance ${\cal R}$ during transmission of the pulse, as well
as after its transmission. The spatial dispersion affects the
transmittance ${\cal T}$ and absorbance
 ${\cal A}$ after passing through a quantum well, when
these values are small.

In Fig. 3 and 4, the optical characteristics are represented at
 $\gamma=0$ and a long stimulating pulse
 $(\gamma_{\,\ell}\ll\Delta \omega,$ Fig.3)
and a pulse of a middle duration, when
$\gamma_{\,\ell}\simeq\Delta \omega$ (Fig.4). Since, in that case,
the real absorption is absent, one have to accept the function
${\cal A}$, defined in (\ref{eq31}), as an energy part, stored up
by a quantum well for the time being due to the interband
transitions  (if ${\cal A}>0$), or an energy part, which is
generated by the quantum well during and after propagation of the
pulse (${\cal A}<0$). The same concerns to Fig.2, however, the
part of the stored energy there, which disappears  if  $\gamma\to
0$, corresponds to the real absorption.
 \begin{figure}
\includegraphics [] {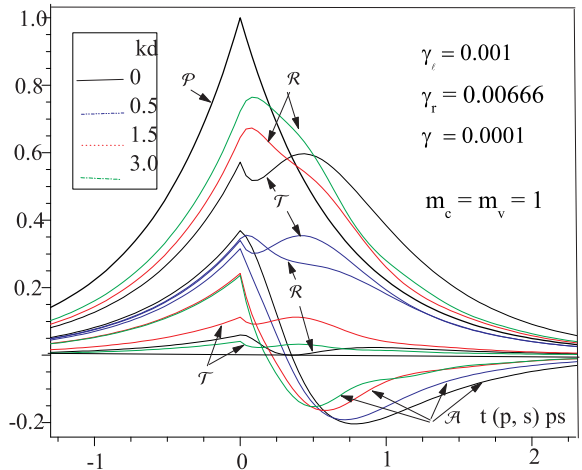} 
\caption[*]{\label{Fig5.eps} Same as in Fig.1 for four magnitudes
of the parameter $kd$ in the case, when $\Delta\omega$ is close to
$\gamma_{\,r}, \gamma_{\,\ell\,}\gg\gamma_{\,r}\gg\gamma.$}
\end{figure}
The oscillation period in Fig.2 and 4 does not depend on the
parameter $kd$, since, at chosen magnitudes of the parameters
$\Delta\omega$ and $\gamma_{\,r}$ the beating frequency
(\ref{eq36}) is almost equal to $\omega_1-\omega_2$, and
comparatively small changes of the functions $\varepsilon'$ and
$\varepsilon''$ does not affect practically on the beating
frequency.

In Fig.5, where $\Delta\omega$ is near $\gamma_{\,r}$
($6.65.10^{-3}eV$ and $6.66.10^{-3}eV$, respectively), the
stimulating light pulse is 5 times shorter, than in Fig.3 and 4,
and $\gamma_{\,\ell}\gg\gamma_{\,r}\gg\gamma$. In that case, the
spatial dispersion affects strongly on the optical
characteristics. In the interval $0\leq kd\leq 3$, the reflectance
increases  8 times approximately, and the transmittance decreases
6 times. Such a sharp change is due to the dependence of
$\bar{\gamma}_{\,r1}$ and $\bar{\gamma}_{\,r2}$ on $kd$. For
example, at $kd=0$ $\bar{\gamma}_{\,r1}=-17303.9,
\bar{\gamma}_{\,r2}=193066,6$, and at  $kd=3$
$Re\,\bar{\gamma}_{\,r1}=1960.21,
Re\,\bar{\gamma}_{\,r2}=442,718$. And at the same time, ${\cal R}$
and ${\cal T}~~\leq 1$, since they are the result of substraction
of large magnitudes and therefore these differences are sensitive
to changes of $kd$.

In \cite{bb14}, it was shown that, at $kd=0$, there are the
singular points on the time axis, where ${\cal T}={\cal A}=0$ and
${\cal R}={\cal P}$, or ${\cal R}={\cal A}=0$ and ${\cal T}={\cal
P}$ (total reflection or total transmission). It is seen from the
figures that the singular points are preserved and in the case
$kd\neq 0$, there is only a small shift of them. In Fig.5 the
point of the total transmission appears at $kd=0$. At $kd= 0.5$,
this point disappears, and at $kd=1.5$ and $kd=3.0$ the point of
the total reflection appears. If $kd=1.5$, then  ${\cal R}={\cal
P}$, ${\cal A}+{\cal T}=0~~ ({\cal A}<0)$. If $kd=3.0$, then as
before ${\cal R}={\cal P}$, but ${\cal A}={\cal T}=0$. Thus,
growing of the parameter  $kd$ changes the type of a singular
point.

Thus, the spatial dispersion of the electromagnetic waves, forming
the light pulse, noticeably  affect the optical characteristics of
a quantum well. This influence is especially strong, when
$\gamma_{\,r}\simeq\Delta\omega$.

Let us note in conclusion that the results obtained above are
valid at equal refraction indices of barriers and quantum well.
Otherwise, one is to take into account reflection of boundaries of
a quantum well. However, this problem is outside the scopes of
present article.

\end{document}